\documentclass[multphys,vecphys]{svmult}

\usepackage{makeidx}         
\usepackage{graphicx}        
\usepackage{multicol}        
\usepackage[bottom]{footmisc}

\makeindex             

\begin{document}

\title*{Testing the BH 176 and Berkeley 29 Association with GASS/Monoceros}
\titlerunning{Orbit of Be29 and GASS?} 
\author{Peter M.~Frinchaboy}
\institute{Univ. of Virginia, Dept. of Astronomy, \\ P.O. Box 400325, \\Charlottesville, VA 22904-4325
\texttt{pmf8b@virginia.edu}}
%
%
\maketitle
It has been previously noted that the outermost open clusters in the Milky Way seem to
lie in a string-like configuration that is coincident with, and may be associated to, the Galactic
anticenter stellar structure (GASS) or the ``Monoceros Ring'' \cite{frinchaboy:martin04,frinchaboy:pmf04}.
Among the clusters that have been suggested to be associated with GASS are Berkeley 29 (Be29) and BH176, 
which have recently had their proper motion ($\mu$) determined by \cite{frinchaboy:dias06}(hereafter D06).
Matching the $\mu$ determinations from D06, to previously published radial velocities (RVs) for 
Berkeley 29 \cite{frinchaboy:pmf06,frinchaboy:carraro04} and BH 176 \cite{frinchaboy:pmf06} 
allows an attempt to derive their orbits.

D06 provided positions and 2MASS photometry \cite{frinchaboy:2mass} for stars in the fields of 
hundreds of open clusters derived from Tycho-2 and UCAC-2 $\mu$, including BH176 and Be29.  
We matched the 2MASS photometry and positions from D06 to the photometry and RV samples for Be29 and BH176 
from F06.   
We find that the stars used by D06 to determine a bulk $\mu$ for BH 176 are too bright to be cluster members, 
as shown in the ($V$,$V-I$) color-magnitude diagram (CMD, Fig.~1a) from \cite{frinchaboy:pmf06}.  
For Be29, we find a similar problem for most of the D06 $\mu$ stars (Fig.~1b); however, 
the tip of the Be29 red giant branch does overlap the faint red end of the UCAC-2 sample.  
Comparing the two samples
finds only one star in common (i.e., having both a measured $\mu$ and an RV consistent with cluster membership), 
but this does allow exploration of a possible orbit for Be29.

The orbital motions of Be29 were calculated by using an orbit integrator and Galactic model 
from \cite{frinchaboy:jsh95}. The orbit is followed backwards for a time interval equal 
to the age of each cluster (3.7 Gyr; \cite{frinchaboy:pmf06}). We find that the resulting orbit, shown in 
Figure 2, is roughly consistent with the orbit for GASS \cite{frinchaboy:penn05}. 
%
The preliminary orbit for Be29 shows that the cluster and GASS share a similar tilt 
with respect to the Galactic plane ($\sim 40 \pm 30 \deg$ for Be29 and $\sim 17 \deg$ for GASS;\cite{frinchaboy:pmf06}),
though of course further work is needed to improve the cluster motion. Moreover, whether 
GASS is an accretion product of just a flare of the disk is still debated\cite{frinchaboy:momany06}.   
A detailed 3-D kinematic follow-up on possible GASS 
clusters is needed to determine whether their origins due to are normal formation or accretion.
\vskip0.05in
PMF acknowledges funding by the F.H. Levinson 
Fund of the Peninsula Community Foundation, a  
NASA GSRP, U. Virginia 
Faculty Senate Dissertation Fellowship, the Virginia Space Grant Consortium, 
and by a AAS-ITG.
\vskip-0.2in
%
%
%
%
%
%
%
%
\begin{figure}
\centering
\includegraphics[height=2.15in]{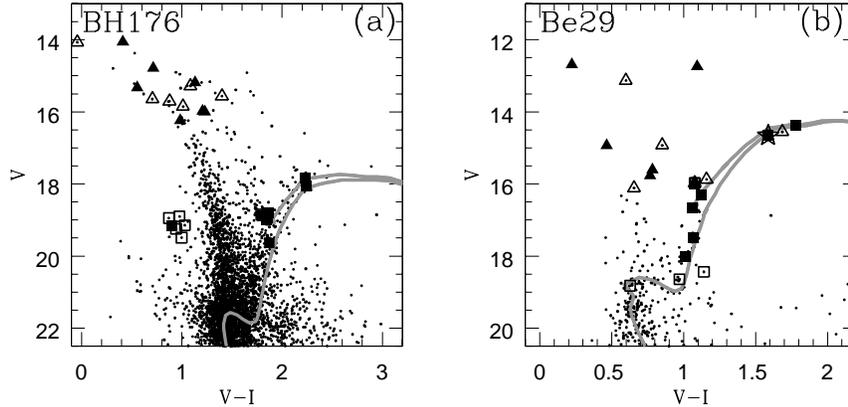}
%
%
\vskip-0.12in
\caption{a) Color-magnitude diagram (CMD) for BH176 from F06. 
Triangles denote proper motion stars from D06, with filled triangles being stars that have probabilities $> 60$\%. 
Squares are RV stars from F06, with filled squares denoting F06 RV members and open squares the non-members. 
The \cite{frinchaboy:girardi00} isochrone match from F06 is overplotted in grey.  b) Same as a) for Be29.  
The star denotes star used for orbit. }
\label{fig:1}       
\end{figure}
\begin{figure}
\centering
\vskip-0.13in
\includegraphics[width=4.3in]{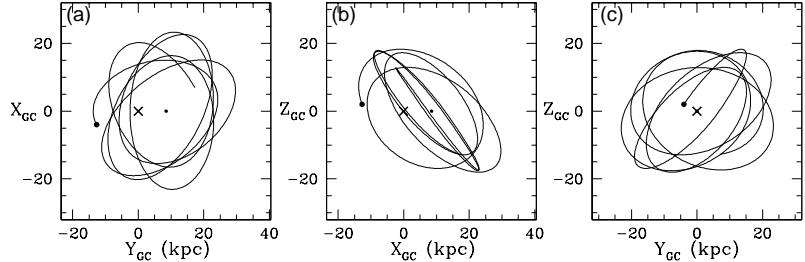}
%
%
\vskip-0.17in
\caption{Orbit for Be29 based on the one star in common between the RV membership of F06 and 
with proper motion from D06. The large dot denotes the cluster's current position, while the 
cross denotes the position of the Sun. This preliminary Be29 orbit shows the same orientation 
as GASS\cite{frinchaboy:penn05} in the Galactic $X,Y$ plane (b).}
\label{fig:2}       
\vskip-0.15in
\end{figure}
%
%
%
%
%
%
\vskip-0.25in
%
%

%
%

\vskip-0.25in
\printindex
\end{document}